\documentclass[twocolumn,prl,showpacs,preprintnumbers,amsmath,amssymb,superscriptaddress]{revtex4}
\usepackage{graphicx}
\usepackage{dcolumn}
\usepackage{bm}

\font\scripti=cmmi7
\font\scriptscripti=cmmi5
\def\sib#1{\setbox0 = \hbox{\scripti #1}
  \kern-.02em\copy0\kern-\wd0
  \kern.04em\box0} 
\def\ssib#1{\setbox0 = \hbox{\scriptscripti #1}
  \kern-.02em\copy0\kern-\wd0
  \kern.04em\box0} 
\font\tenib=cmmib10 
\skewchar\tenib='177 \skewchar\tenib='177 \skewchar\tenib='177
\def\pbold#1{\setbox0 = \hbox{$ #1 $}
  \kern-.022em\copy0\kern-\wd0
  \kern.011em\copy0\kern-\wd0
  \kern.011em\copy0\kern-\wd0
  \kern.011em\copy0\kern-\wd0
  \kern.011em\box0} 

\usepackage{graphicx}
\usepackage{dcolumn}
\usepackage{bm}
\usepackage{color}

\def\up{\uparrow}
\def\dwn{\downarrow}

\def\lesssim{\ \raise.3ex\hbox{$<$}\kern-0.8em\lower.7ex\hbox{$\sim$}\ }
\def\gesim{\ \raise.3ex\hbox{$>$}\kern-0.8em\lower.7ex\hbox{$\sim$}\ }

\begin{document}
\title{Low-Dimensional Fluctuations and Pseudogap in Gaudin-Yang Fermi Gases}
\author{Hiroyuki Tajima}
\affiliation{Department of Mathematics and Physics, Kochi University, Kochi 780-8520, Japan}
\author{Shoichiro Tsutsui}
\affiliation{Quantum Hadron Physics Laboratory, RIKEN Nishina Center, Wako, Saitama, 351-0198, Japan}
\author{Takahiro M. Doi}
\affiliation{Research Center for Nuclear Physics (RCNP), Osaka University, Osaka 567-0047, Japan}
\date{\today}
\begin{abstract}
Pseudogap is a ubiquitous phenomenon in strongly correlated systems such as high-$T_{\rm c}$ superconductors, ultracold atoms and nuclear physics. 
While pairing fluctuations inducing the pseudogap are known to be enhanced in low-dimensional systems, such effects have not been explored well
in one of the most fundamental 1D models, that is, Gaudin-Yang model.
In this work, we show that the pseudogap effect can be visible in the single-particle excitation in this system using a diagrammatic approach.
Fermionic single-particle spectra exhibit a unique crossover from the double-particle dispersion to pseudogap state with increasing the attractive interaction and the number density at finite temperature.
Surprisingly, our results of thermodynamic quantities in unpolarized and polarized gases show an excellent agreement with the recent quantum Monte Carlo and complex Langevin results, even in the region where the pseudogap appears. 
\end{abstract}
\maketitle
A pseudogap phenomenon, which is the suppression of the density of states (DOS) around a Fermi level, has been a central issue in strongly-correlated quantum many-body systems such as high-$T_{\rm c}$ superconductors~\cite{Timusk,Fischer,Yanase,Varma,Hashimoto}, ultracold atoms~\cite{Chen,Bloch,Giorgini,Zwerger,Randeria,Mueller,Strinati,Jensen,Ohashi}, and nuclear and quark matter~\cite{Schnell,Kitazawa,Kitazawa2,Abe,Huang,Pang}.
While the origin of the pseudogap strongly depends on the properties of each system,
it is believed that the pseudogap is induced by fluctuation effects dominating nontrivial characters of the systems.
Recently, an ultracold atomic gas provides us an ideal platform to study the pseudogap physics and associated fluctuation effects in a systematic way~\cite{Haussmann,Chen2,Tsuchiya,Hu,Perali2,Magierski,Su,Stewart,Gaebler,Sagi,Wlazlowski,Jensen2,Halford}, thanks to the realization of the Bardeen-Cooper-Schrieffer (BCS) to Bose-Einstein-Condensation (BEC) crossover~\cite{Eagles,Leggett,NSR,SadeMelo,OhashiGriffin,Regal,Zwierlein}. 
\par
Furthermore, the low-dimensionality tends to induce strong fluctuations~\cite{Lee,Luther}.
It is a key point also for properties of carbon nanotubes~\cite{Dresselhaus,Meunier}, organic conductors~\cite{Little,Heeger,Jerome}, as well as nuclear {\it pasta} in neutron star crusts~\cite{Caplan,Iida}.
In ultracold atom physics, two-dimensional pseudogap effects have attracted much attention~\cite{Feld,Frohlich,Pietila,Klimin,Ngampruetikorn,Watanabe2,Marsiglio,Bauer,Matsumoto,Murthy,Mulkerin} because these many-body effects are expected to be more visible than 3D systems.
Along this direction, the pseudogap in a one-dimensional cold atomic system would be a fascinating topic.
\par
While an attractively interacting two-component Fermi gas in 1D, namely, Gaudin-Yang model is known as a solvable model based on the thermodynamic Bethe ansatz (TBA)~\cite{Guan},
it does not mean that physical quantities we are interested in can easily be obtained in an exact way.
Low-energy effective field theory descriptions such as Tomonaga-Luttinger liquid (TLL)~\cite{Imambekov} have also been employed frequently in 1D.
While such approaches also give exact results at zero temperature,
it is not the case at finite temperature where the Fermi step is softened.
Indeed, precise results of this 1D fermionic system at finite temperature was not obtained before a recent state-of-the-art work of quantum Monte Carlo (QMC) simulation done by Hoffman, {\it et al.}~\cite{DrutQMC}.
Afterwards, various thermal properties of this system have been investigated within a lattice simulation~\cite{Drutimagmu,Rammelmuller2017,Shill,Alexandru}.
However, no one shows how the pseudogap phenomena occur in this famous 1D model.
Moreover, the possibility of an inhomogeneous pairing state called Fulde-Ferrel-Larkin-Ovchinikov (FFLO)-like state~\cite{FF,LO} has also been extensively investigated in a spin-imbalanced 1D system~\cite{Hu2007,Parish2007,Orso,Zhao,Reza,Liao,Rammelmuller2020} since the 1D FFLO-like state is expected to be robust against fluctuations.
To see this, a quantitative analysis of fluctuation effects are really desired. 
It involves interdisciplinary interests from other fields.
In the context of quantum chromodynamics (QCD), 
the FFLO-like state of quark-antiquark pairs called chiral spiral is anticipated at finite density~\cite{Nakano,Fukushima,Buballa}.
\par
In this work, we elucidate pairing fluctuation effects in 1D Gaudin-Yang Fermi gas at finite temperature within the diagrammatic approach,
which has successfully been applied to higher-dimensional systems~\cite{Chen,Strinati,Ohashi,Zwerger,Mueller}.
Our numerical results of the number density show an excellent agreement with the recent QMC results~\cite{DrutQMC}.
In the polarized case, we show that our result also well reproduces a complex Langevin (CL) simulation~\cite{DrutCL}, which is a promising candidate for overcoming a sign problem in an imbalanced Fermi gas~\cite{Berger}. 
Furthermore, we show that the single-particle excitation spectra exhibit the pseudogapped structure due to pairing fluctuations in the region where the validity is guaranteed by the comparison with QMC results for the thermodynamic quantity. 
\par
We start from the attractive Gaudin-Yang model described by the Hamiltonian
\begin{align}
H&=\sum_{p,\sigma}\xi_{p,\sigma}c_{p,\sigma}^\dag c_{p,\sigma}\cr
&+g\sum_{k,k',q}c_{k+\frac{q}{2},\up}^\dag c_{-k+\frac{q}{2},\dwn}^\dag c_{-k'+\frac{q}{2},\dwn}c_{k'+\frac{q}{2},\up},
\end{align}
where $\xi_{p.\sigma}=p^2/(2m_\sigma)-\mu_{\sigma}$ is the kinetic energy of a fermion with momentum $p$, spin $\sigma=\up,\dwn$ and mass $m_{\sigma}$ measured from the chemical potential $\mu_{\sigma}$.
For simplicity, we consider the mass-balanced case ($m\equiv m_{\up}=m_{\dwn}$).
$\mu_{\sigma}$ is parametrized by the averaged one $\mu=(\mu_{\up}+\mu_{\dwn})/2$ and a fictitious magnetic field $h=(\mu_{\up}-\mu_{\dwn})/2$.
$c_{p,\sigma}$ and $c_{p,\sigma}^{\dag}$ are fermionic annihilation/creation operators, respectively.
The coupling constant $g$ is related to a 1D scattering length $a$ as $g=-\frac{2}{ma}$. 
Following Ref.~\cite{DrutQMC}, we measure the interaction strength through the dimensionless parameter $\lambda^2=m g^2/T$. 
Since the two-body bound state with the binding energy $E_{\rm b}=1/(ma^2)$ always exists in an attractive 1D system,
$\lambda^2=4E_{\rm b}/T$ characterizes the ratio between $E_{\rm b}$ and $T$. 
\par
\begin{figure}[t]
\begin{center}
\includegraphics[width=7cm]{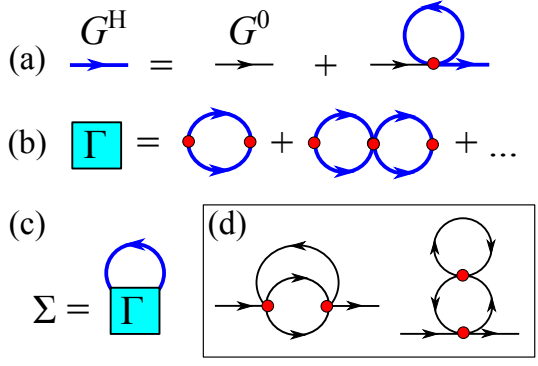}
\end{center}
\caption{Feynman diagrams for (a) Hartree Green's function $G^{\rm H}$ (solid), (b) four-point vertex $\Gamma$, (c) the self-energy $\Sigma$, and (d) the second-order connected diagrams taken into account in our approach. The thin line denotes the bare Green's function $G^0$.
}
\label{fig1}
\end{figure}
The important notice is that while in 2D and 3D systems the contact-type interaction becomes zero such that a finite scattering length is reproduced~\cite{WernerCastin}, it is not the case in this 1D system.
Therefore, the lowest-order diagram, that is, Hartree self-energy $\Sigma_{\sigma}^{\rm H}$ is nonzero, in contrast to higher-dimensional systems.
To retain it,  as shown in Fig.~\ref{fig1} (a) we introduce the single-particle Green's function $G_{\sigma}^{\rm H}(p,i\omega_n)=G_{\sigma}^0(p,i\omega_n)\left[1+\Sigma_{\sigma}^{\rm H}G_{\sigma}^{\rm H}(p,i\omega_n)\right]$ with the Hartree shift $\Sigma_{\sigma}^{\rm H}=gn_{-\sigma}$ [$\omega_n=(2n+1)\pi T$ is the fermion Matsubara frequency and $-\sigma$ represents an opposite spin for $\sigma$] where $G_{\sigma}^0(p,i\omega_n)=[i\omega_n-\xi_{p,\sigma}]^{-1}$ is a bare Green's function.
A similar approximation has been employed in nuclear physics with finite-range interactions~\cite{Jin,Ramanan,Pieter,Tajima2019}.
On the basis of $G_{\sigma}^{\rm H}(p,i\omega_n)$, we incorporate pairing fluctuation effects described by the four-point vertex $\Gamma$ diagrammatically shown in Fig.~\ref{fig1} (b), which reads
\begin{align}
\Gamma(q,i\nu_\ell)=-\frac{g^2\Pi(q,i\nu_\ell)}{1+g\Pi(q,i\nu_\ell)},
\end{align}
where
\begin{align} 
\Pi(q,i\nu_\ell)&=T\sum_{k,i\omega_n}G_{\up}^{\rm H}\left(k+\frac{q}{2},i\nu_{\ell}+i\omega_n\right)\cr
&\quad\quad\times G_{\dwn}^{\rm H}\left(-k+\frac{q}{2},-i\omega_n\right),
\end{align}
is the lowest-order particle-particle bubble with the boson Matsubara frequency $\nu_\ell=2\ell \pi T$.
The self-energy $\Sigma_{\sigma}$ for the fluctuation correction is given by
\begin{align}
\label{eq:sigma}
\Sigma_{\sigma}(p,i\omega_n)=T\sum_{q,i\nu_\ell}\Gamma(q,i\nu_\ell)G_{-\sigma}^{\rm H}(q-p,i\nu_\ell-i\omega_n).
\end{align}
We note that this approximation is equivalent to the so-called $T$-matrix approach, except for the self-consistent treatment of the Hartree shift.
The $T$-matrix approach successfully reproduces the exact results obtained by TBA for 1D Fermi polaronic excitations realized in spin-polarized limit~\cite{Doggen}.
In our approach, by taking $G^{\rm H}(p,i\omega_n)$ with density mean-field $\Sigma_{\sigma}^{\rm H}$ as a building block of fluctuation corrections,
at least we take all possible connected diagrams into account up to the second-order shown in Fig.~\ref{fig1} (d).  
Using the dressed Green's function $G_{\sigma}(p,i\omega_n)=G_{\sigma}^{\rm H}(p,i\omega_n)\left[1+\Sigma_{\sigma}(p,i\omega_n)G_{\sigma}(p,i\omega_n)\right]$,
we obtain the number density $n_\sigma$ for given $T$ and $\mu_{\sigma}$ as
$n_{\sigma}=T\sum_{p,i\omega_n}G_{\sigma}(p,i\omega_n).$
Moreover, we can obtain the single-particle spectral function $A_{\sigma}(p,\omega)=-\frac{1}{\pi}{\rm Im}G_{\sigma}(p,i\omega_n\rightarrow\omega+i\delta)$ and the density of states (DOS) $\rho_\sigma(\omega)=\sum_{p}A_{\sigma}(p,\omega)$ from $G_{\sigma}(p,i\omega_n)$.
In what follows, we suppress $\sigma$ in these quantities as $n$, $A(p,\omega)$, and $\rho(\omega)$ unless otherwise specified. 
\par
\begin{figure}[t]
\begin{center}
\includegraphics[width=7cm]{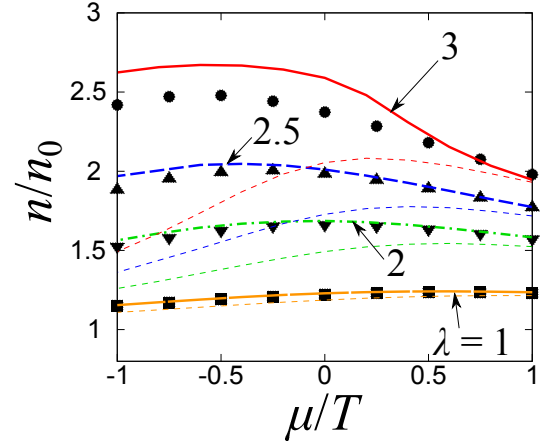}
\end{center}
\caption{Calculated number density $n/n_0$ as function of $\mu/T$ in an unpolarized system,
where $n_0$ is the non-interacting counterpart.
The thick and thin curves show the numerical results of our diagrammatic approach and the Hartree-Fock calculation with $\Sigma_{\sigma}^{\rm H}$, respectively.
The interaction parameter $\lambda$ is given by $1$, $2$, $2.5$, and $3$ from the bottom.
The black symbols represents 1D QMC results~\cite{DrutQMC} of $\lambda=1$ (square), $2$ (inverted triangle), $2.5$ (triangle), and $3$ (circle).
}
\label{fig2}
\end{figure}
Figure~\ref{fig2} shows the calculated number density $n/n_0$ as a function of $\mu/T$ in an unpolarized gas, where $n_0$ is the non-interacting counterpart.
Our results given by thick curves show an excellent agreement with 1D QMC results from Ref.~\cite{DrutQMC}.
For comparison, we also plot the Hartree-Fock results (thin curves) given by $n^{\rm H}=T\sum_{p,i\omega_n}G^{\rm H}(p,i\omega_n)$.
While all results coincides with each other in the weak-coupling regime such as $\lambda=1$,
the Hartree-Fock result deviates from the others due to the lack of fluctuation effects.
Our main results well reproduce the QMC results even in the strong-coupling regime ($\lambda\geq2$) where $E_{\rm b}=\lambda^2T/4\geq T$.
While the result at $\lambda=3$ is close to the applicable limit of our approach as we will mention later, still it shows a semi-quantitative agreement with QMC.
In this way, we can check the validity of our approach in these parameter regimes.
We note that these results are also consistent with TBA~\cite{He}. 
\par
\begin{figure}[t]
\begin{center}
\includegraphics[width=\hsize]{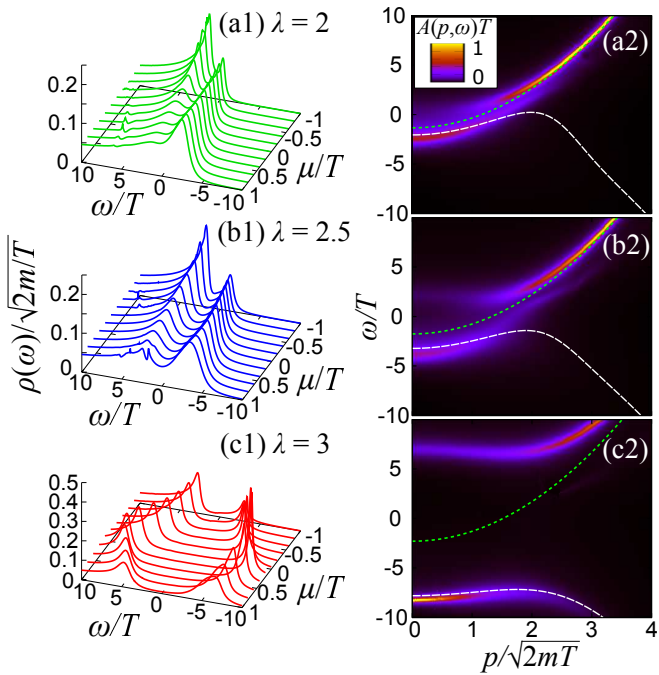}
\end{center}
\caption{Calculated DOS $\rho(\omega)$ with $h=0$ at (a1) $\lambda=2$, (b1) $2.5$, and (c1) $3$. The panels (a2), (b2), and (c2) show the corresponding single-particle spectral functions $A(p,\omega)$ at $\mu/T=0.4$.
In right panels, the dashed and dotted curves represent $\xi_{p,\sigma}^{\rm eff}$ obtained from Eq. (\ref{eq:xieff}) with $G$ and $G^{\rm H}$, respectively. 
}
\label{fig3}
\end{figure}
Figures~\ref{fig3} (a1), (b1), and (c1) show the calculated DOS $\rho(\omega)$ at $\lambda=2$, $2.5$, and $3$.
One can see the dip structure around $\omega=0$ (corresponding to the Fermi level) with the double peaks in the wide parameter region.
While the higher-energy peak locates at $\omega=-\mu$ in the low-density regime ($\mu/T\lesssim 0$), the energy of lower one is approximately given by $\omega=-E_{\rm b}/2$. This indicates the existence of two-body bound molecules.
With increasing the density (in other words, $\mu/T$ or $\lambda$), one can find pronounced gap structure even in the high-density regime ($\mu/T\gesim 0$).
It is expected to originate from many-body effects, namely, pairing fluctuations associated with the Cooper instability.
\par
While the dip structures in Fig.~\ref{fig3} are similar to the pseudogap,
we have to carefully distinguish the fluctuation-induced pseudogap and the double peak due to the two-body bound state.
For this purpose, $A(p,\omega)$ is useful.
At an intermediate coupling ($\lambda=2$) in Fig.~\ref{fig3} (a2) $A(p,\omega)$ is largely broadened around $\omega=0$.
While the obtained spectra is somewhat similar to those in the Luther-Emery model at $T=0$~\cite{Voit,Orignac},
the broadening and renormalization of the dispersion through the self-energy corrections are significant even in the relatively high-energy region ($|\omega|\gesim T$). 
The double-particle (two quadratic) dispersions in Fig.~\ref{fig3} (a2) can be qualitatively understood from the coupling between dressed atoms and thermally excited molecules at finite momenta $q$, which is characterized by $\Gamma(q,i\nu_\ell)$ in Eq.~(\ref{eq:sigma}).
While a similar spectrum can be found in the BEC regime in higher dimensions~\cite{Tsuchiya,Watanabe2},
such a strong intensity of the double-particle dispersion is a remarkable feature of this 1D model. 
With increasing the interaction strength as shown in Fig.~\ref{fig3} (b2), the two dispersions are separated, and a hole-like contribution appears at positive energy.
The overall structure gradually changes into the pseudogapped spectrum.
At stronger coupling in Fig.~\ref{fig3} (c2),
since the low-energy pole in $\Gamma(q,i\nu_\ell)$ becomes close to $q=0$ and gives a strong particle-hole coupling, one can clearly see the pseudogap accompanying with particle-hole branches.
Intuitively, this pseudogap structure can be understood from the so-called static approximation where $\Sigma_{\sigma}^{\rm pg}(p,i\omega_n)\simeq-\Delta_{\rm pg}^2G_{-\sigma}^{\rm H}(-p,-i\omega_n)$~\cite{Strinati,Ohashi}. 
Here, $\Delta_{\rm pg}^2=-T\sum_{q,i\nu_\ell}\Gamma(q,i\nu_\ell)$ is called the pseudogap parameter which characterizes its size in $A(p,\omega)$ as well as $\rho(\omega)$.
Indeed, this approximated self-energy induces the BCS-like gapped DOS. 
These results indicate the crossover from the superposition of atoms and diatomic pairs to the pseudogap state with increasing the interaction at finite temperature.
The pseudogapped dispersion of fermions yields that the elementary excitation is now replaced by the bosonic two-particle excitations~\cite{Fuchs}.
We note that in this work we do not specify the crossover boundary between the pseudogap regime and the bound molecular regime since its definition involves ambiguity~\cite{Mueller,Strinati,Ohashi}. 
\par
We also compare the dispersion $\xi_{p,\sigma}^{\rm eff}$ obtained from the imaginary-time Green's function $G_{\sigma}(p,\tau)$ as
\begin{align}
\label{eq:xieff}
\xi_{p,\sigma}^{\rm eff}=\frac{1}{\Delta \tau}\ln\left|\frac{G_{\sigma}(p,\tau+\Delta\tau)}{G_{\sigma}(p,\tau)}\right|_{\tau\rightarrow \beta},
\end{align}
where $\Delta\tau$ is a small number and $\beta=1/T$ is the inverse temperature.
We take $\Delta\tau=\beta/40$, which is enough small to extract $\xi_{p,\sigma}^{\rm eff}$.
We note that a large $\tau$ limit ($\tau\rightarrow \beta$) is required to obtain the ground-state single-particle energy.
Such an extraction of the dispersion has frequently been done to obtain hadronic spectra in lattice QCD simulations~\cite{Rothe}.
Indeed, in the single-particle case in vacuum with $G_{\sigma}(p,\tau)\propto e^{-\frac{p^2}{2m}\tau}$, one can obtain $\xi_{p,\sigma}^{\rm eff}=p^2/(2m)$ from Eq. (\ref{eq:xieff}).
In the present case with strongly correlated media,
while at small momenta $\xi_{p,\sigma}^{\rm eff}$ well reproduces the peak in $A_{\sigma}(p,\omega)$ (see dashed curves in the right panels in Fig.~\ref{fig3}), it deviates from the peak at weaker coupling side due to the broadening of spectra as well as level couplings at high momenta.
In such a high-energy regime where interaction effects are irrelevant,
the dispersion obtained from $G_{\sigma}^{\rm H}(p,\tau)$ agrees with the spectral peak.
On the other hand, in the deep inside of the pseudogap regime such as Fig.~\ref{fig3} (c2), 
$\xi_{p,\sigma}^{\rm eff}$ shows a good agreement with the so-called back-bending curve in $A(p,\omega)$.
We note that this quantity can be measured in the lattice simulation without analytic continuations.
The medium corrections on $\xi_{p,\sigma}^{\rm eff}$ in many-body systems would be useful information for the future investigation of finite-density lattice QCD simulations.
\begin{figure}[t]
\begin{center}
\includegraphics[width=7cm]{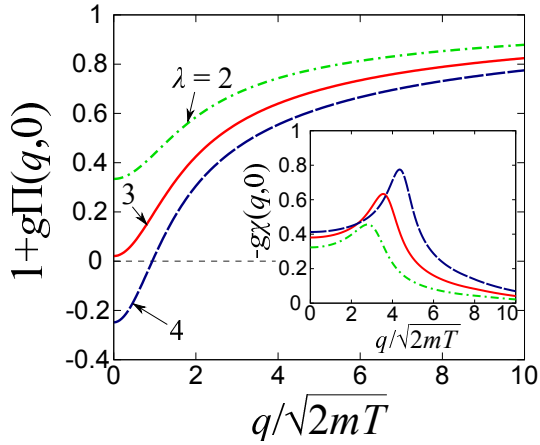}
\end{center}
\caption{The denominator of $\Gamma(q,0)$ given by $1+g\Pi(q,0)$ at $\beta\mu=1$. 
The inset shows the Lindhard function $-g\chi(q,0)$ at finite temperature.
}
\label{fig4}
\end{figure}
\par
\begin{figure}[t]
\begin{center}
\includegraphics[width=7cm]{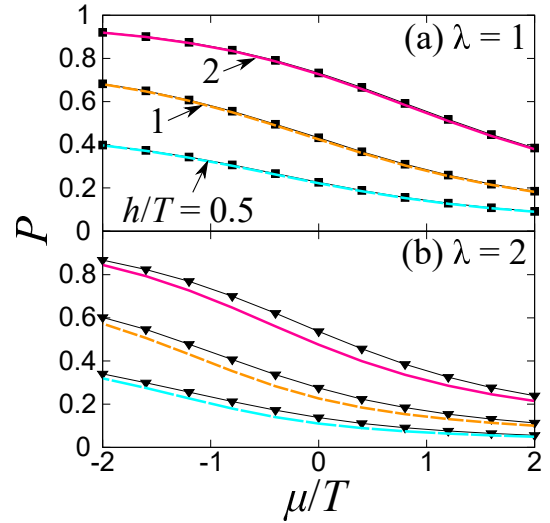}
\end{center}
\caption{The polarization equation of state $P=(n_\up-n_\dwn)/(n_\up+n_\dwn)$ at (a) $\lambda=1$ and (b) $\lambda=2$. 
The calculated $P$ in our diagrammatic approach are plotted at $h/T=0.5$, $1$, and $2$ from the bottom to the top in each figure.
The filled symbols show the numerical results of the CL method~\cite{DrutCL}.
}
\label{fig5}
\end{figure}

We note that our diagrammatic approach has an artificial limitation in the strong-coupling regime. 
Fig.~\ref{fig5} shows the denominator of $\Gamma(q,0)$.
While in 3D systems the superfluid transition is identified by $1+g\Pi(0,0)=0$,
it should be positive due to the Mermin-Wagner-Hohenberg theorem~\cite{Mermin,Hohenberg} yielding no phase transition in uniform 1D systems.
On the other hand, we encounter the zero-crossing of $\Gamma(q,0)$ around $\lambda\gesim 3.1$ due to the lack of higher-fluctuation corrections.
However, we emphasize that our results show non-trivial spectral structures
even in the region where our approach is valid and the calculated number density quantitatively agrees with the QMC results. 
\par
In 1D systems, Peierls instability may also occur through the density response function $\chi(q,i\nu_\ell)$ given by
\begin{align}
\chi(q,i\nu_\ell)=-\sum_{k}\frac{f(\xi_{k-q,\sigma}^{\rm H})-f(\xi_{k,\sigma}^{\rm H})}{i\nu_\ell+\xi_{k-q,\sigma}^{\rm H}-\xi_{k,\sigma}^{\rm H}}.
\end{align}
It is nothing but the Lindhard function~\cite{FW}, which is known to show the logarithmic divergence with respect to $T$ at $q=2k_{\mu}\equiv2\sqrt{2m(\mu-\Sigma^{\rm H})}$ and $\nu_\ell=0$ in 1D.
$\chi(q,i\nu_\ell)$ is involved in the second-order self-energy diagram $\Sigma_{\sigma}^{\rm 2nd}(p,i\omega_n)$ in our approach as
\begin{align}
\Sigma_{\sigma}^{\rm 2nd}(p,i\omega_n)&=g^2T\sum_{q,i\nu_\ell}\chi(q,i\nu_\ell)\cr
&\quad\times G_{\sigma}^{\rm H}(p-q,i\omega_n-i\nu_\ell),
\end{align}
which is topologically equivalent to the first diagram in Fig.~\ref{fig1}(d) with replacing $G^0$ with $G^{\rm H}$.
If $\chi(q,0)$ has such a divergence, one can also obtain the approximate self-energy inducing the Peierls pseudogap $\Delta_{\rm Pi.}$~\cite{Lee,Bartosch} as $\Sigma_{\sigma}^{\rm 2nd}(p,i\omega_n)\simeq-\Delta_{\rm Pi.}^{2}G_{\sigma}^{\rm H}(p\pm 2k_{\mu},i\omega_n)$. 
The inset of Fig.~\ref{fig4} shows the calculated $-g\chi(q,0)$ at $\mu/T=1$.
Although $-g\chi(q,0)$ exhibits a maximum around $q=2k_{\mu}$,
it is still finite due to the finite temperature effect.
Such a softening of the anomaly in $\chi(q,0)$ is one of the reasons why our diagrammatic approach unexpectedly well reproduces the QMC results at finite temperature.
Even in the analysis based on the random phase approximation for the density channel, since the Peierls instability is identified by $1+g\chi(q,0)=0$, the region where we explore in this work is safely far away from this instability.
If one incorporates higher-order density fluctuation effects in more sophisticated approaches such as fluctuation-exchange approximation~\cite{Yanase}, one may expect the competition of two pseudogaps originating from Cooper and Peierls instabilities even in this simple model, which is left as interesting future work. 
\par
Finally, in Fig.~\ref{fig5} we have also plotted the polarization $P=(n_\up-n_\dwn)/(n_\up+n_\dwn)$ in the presence of finite fictitious magnetic field $h/T$.
We compare our results with the CL method~\cite{DrutCL} which is developed to avoid a possible sign problem in polarized systems.
We note that the CL method also agrees with other methods such as lattice simulation with an analytic continuation from the imaginary chemical potential~\cite{Drutimagmu,Alexandru,DrutCL}.
Our results well reproduce the CL results, indicating that our approach enables us to evaluate fluctuation effects quantitatively even in the presence of polarization.
Thus one can expect possible future applications of our diagrammatic approach to other interesting problems such as fluctuation effects on FFLO-like pairing states, transport in quantum wires~\cite{Giamarchi}, and multi-polaronic excitations~\cite{Zinner,Mistakidis}.
\par
In conclusion,
we have investigated low-dimensional fluctuation effects in an attractive Gaudin-Yang Fermi gas at finite temperature within the diagrammatic approach.
The calculated number densities and polarizations in unpolarized and polarized gases shows an excellent agreement with the recent QMC and CL results in the wide ranges of an interaction parameter and chemical potentials.
These results indicate the reliability of our approach in the region where we have explored in this work.
The single-particle spectral functions exhibit the crossover from the superposition of dressed atomic state and thermal dimers to the pseudogap state with increasing the interaction strength and the number density. 
Our analysis can be extended to the mass-imbalanced mixtures and the trapped systems.
It is also interesting to address photo-emission spectra which can be experimentally measured.
\par
The authors thank J. E. Drut and A. C. Loheac for sharing us with their numerical data, 
K. Iida for reading the manuscript and giving insightful comments,
and T. Hatsuda, M. Horikoshi, S. Inoue, K. Kato, and Y. Sekino for useful discussions.
This work is supported by Grants-in-Aid for JSPS fellows (No.17J03975) and for Scientific Research from JSPS (No. 18H05406).


\begin{thebibliography}{99}

\bibitem{Timusk} T. Timusk and B. Statt,
Rep. Prog. Phys. \textbf{62}, 61 (1999).
\bibitem{Yanase} Y. Yanase, T. Jujo, T. Nomura, H. Ikeda, T. Hotta, and K. Yamada,
Phys. Rep. \textbf{387}, 1 (2003).
\bibitem{Fischer} \O. Fischer, M. Kugler, I. Maggio-Aprile, C. Berthod, and C. Renner,
Rev. Mod. Phys. \textbf{79}, 353 (2007).
\bibitem{Varma} C. Varma, 
Nature \textbf{468}, 184 (2010).
\bibitem{Hashimoto} M. Hashimoto, I. M. Vishik, H. He, T. P. Devereaux, and Z.-X. Shen,
Nat. Phys. \textbf{10}, 483 (2014).

\bibitem{Chen} Q. Chen, J. Stajic, S. Tan, and K. Levin,
Phys. Rep. \textbf{412}, 1 (2005).
\bibitem{Bloch} 
I. Bloch, J. Dalibard, and W. Zwerger,
Rev. Mod. Phys. \textbf{80}, 885 (2008).
\bibitem{Giorgini} 
S. Giorgini, L. P. Pitaevskii, and S. Stringari,
Rev. Mod. Phys. \textbf{80}, 1215 (2008).
\bibitem{Zwerger} {\it The BCS-BEC Crossover and the Unitary Fermi Gas },
edited by W. Zwerger, Lecture Notes in Physics Vol. 836 (Springer, Berlin, 2012).
\bibitem{Randeria} M. Randeria and E. Taylor,
Annu. Rev. Condens. Matter Phys. \textbf{5}, 209 (2014).
\bibitem{Mueller} E. J. Mueller,
Rep. Prog. Phys. \textbf{80}, 104401 (2017).
\bibitem{Strinati} G. C. Strinati, P. Pieri, G. R\"{o}pke, P, Schuck, and M. Urban,
Phys. Rep. \textbf{738}, 3 (2018).
\bibitem{Jensen} S. Jensen, C. N. Gilbreth, and Y. Alhassid,
Eur. Phys. J. Spec. Top. \textbf{227}, 2241 (2019).

\bibitem{Ohashi} Y. Ohashi, H. Tajima, and P. van Wyk,
Prog. Part. Nucl. Phys. \textbf{111}, 103739 (2020).

\bibitem{Schnell} A. Schnell, G. R\"{o}pke, and P. Schuck,
Phys. Rev. Lett. \textbf{83}, 1926 (1999).
\bibitem{Kitazawa} M. Kitazawa, T. Koide, T. Kunihiro, and Y. Nemoto,
Phys. Rev. D \textbf{65}, 091504 (2002).
\bibitem{Kitazawa2} M. Kitazawa, T. Kunihiro, and Y. Nemoto,
Phys. Lett. B \textbf{633}, 269 (2006).
\bibitem{Abe} T. Abe and R. Seki,
Phys. Rev. C \textbf{79}, 054002 (2009).
\bibitem{Huang} X.-G. Huang,
Phys. Rev. C \textbf{81}, 034007 (2010).
\bibitem{Pang} J. Pang, J. Wang, and L. He,
Phys. Rev. D \textbf{88}, 054017 (2013).

\bibitem{Stewart} J. T. Stewart, J. P. Gaebler, and D. S. Jin, Nature {\bf 454}, 744 (2008).
\bibitem{Haussmann} R. Haussmann, M. Punk, and W. Zwerger,
Phys. Rev. A \textbf{80}, 063612 (2009).
\bibitem{Chen2} Q. Chen and K. Levin,
Phys. Rev. Lett. \textbf{102}, 190402 (2009).
\bibitem{Tsuchiya} S. Tsuchiya, R. Watanabe, and Y. Ohashi, Phys. Rev. A \textbf{80}, 033613 (2009).
\bibitem{Su} S-Q. Su, D. E. Sheehy, J. Moreno, and M. Jarrell, Phys. Rev. A \textbf{81}, 051604(R) (2010).
\bibitem{Hu} H. Hu, X.-J. Liu, P. D. Drummond, and H. Dong,
Phys. Rev. Lett. \textbf{104}, 240407 (2010).
\bibitem{Gaebler} J. P. Gaebler, J. T. Stewart, T. E. Drake, D. S. Jin, A. Perali, P. Pieri, and G. C. Strinati, Nat. Phys. \textbf{6}, 569 (2010).
\bibitem{Perali2} A. Perali, F. Palestini, P. Pieri, G. C. Strinati, J. T. Stewart, J. P. Gaebler, T. E. Drake, and D. S. Jin, Phys. Rev. Lett. \textbf{106}, 060402 (2011).
\bibitem{Magierski} P. Magierski, G. Wlaz\l owski, and A. Blugac, Phys. Rev. Lett. \textbf{107}, 145304 (2011).
\bibitem{Wlazlowski} G. Wlaz\l owski, P. Magierski, J. E. Drut, A. Bulgac, and K. J. Roche, Phys. Rev. Lett. \textbf{110}, 090401 (2013).

\bibitem{Sagi} Y. Sagi, T. E. Drake, R. Paudel, R. Chapurin, and D. S. Jin,, Phys. Rev. Lett. \textbf{114}, 075301 (2015).
\bibitem{Jensen2} S. Jensen, C. N. Gilbreth, and Y. Alhassid, 
Phys. Rev. Lett. \textbf{124}, 090604 (2020).
\bibitem{Halford} A. Richie-Halford, J. E. Drut, and A. Bulgac,
arXiv:2004.05014


\bibitem{Eagles} D. M. Eagles Phys. Rev. \textbf{186}, 456 (1969).
\bibitem{Leggett} A. J. Leggett, in {\it Modern Trends in the Theory of Condensed Matter}, edited by A. Pekalski and J. Przystawa (Springer Verlag, Berlin, 1980), p. 14.
\bibitem{NSR} P. Nozi{\`e}res and S. Schmitt-Rink, J. Low Temp. Phys. \textbf{59}, 195 (1985).
\bibitem{SadeMelo} C. A. R. Sa de Melo, M. Randeria, and J. R. Engelbrecht, Phys. Rev. Lett. \textbf{71}, 3202 (1993).
\bibitem{OhashiGriffin} Y. Ohashi and A. Griffin,
Phys. Rev. Lett. \textbf{89}, 130402 (2002).
\bibitem{Regal} C. A. Regal, M. Greiner, and D. S. Jin, Phys. Rev. Lett. \textbf{92}, 040403 (2004).
\bibitem{Zwierlein}M. W. Zwierlein, C. A. Stan, C. H. Schunck, S. M. F. Raupach, A. J. 
Kerman, and W. Ketterle, Phys. Rev. Lett. \textbf{92}, 120403 (2004).


\bibitem{Lee} P. A. Lee, T. M. Rice, and P. W. Anderson,
Phys. Rev. Lett. \textbf{31}, 462 (1973).
\bibitem{Luther} A. Luther and I. Peschel,
Phys. Rev. B \textbf{9}, 2911 (1974).

\bibitem{Dresselhaus} M. S. Dresselhaus, G. Dresselhaus, and R. Saito,
Carbon, \textbf{33}, 883 (1995).
\bibitem{Meunier} V. Meunier, A. G. Souza Filho, E. B. Barros, and M. S. Dresselhaus,
Rev. Mod. Phys. \textbf{88}, 025005 (2016).

\bibitem{Little} W. A. Little, Phys. Rev. \textbf{134}, A1416 (1964).
\bibitem{Heeger} A. J. Heeger, S. Kivelson, J. R. Schrieffer, and W.-P. Su.
Rev. Mod. Phys. \textbf{60}, 781 (1988).
\bibitem{Jerome} D. J\'{e}rome, Chem. Phys. \textbf{104}, 5565 (2004).


\bibitem{Iida} G. Watanabe, K. Iida, and K. Sato, Nucl. Phys. A \textbf{676}, 455 (2000).
\bibitem{Caplan} M. E. Caplan and C. J. Horowitz, Rev. Mod. Phys. \textbf{89}, 041002 (2017).

\bibitem{Feld} M. Feld, B. Fr\"{o}hlich, E. Vogt, M. Koschorreck, and M. K\"{o}hl,
Nature \textbf{480}, 75 (2011).
\bibitem{Frohlich} B. Fr\"{o}hlich, M. Feld, E. Vogt, M. Kschorreck, W. Zwerger, and M. K\"{o}hl,
Phys. Rev. Lett. \textbf{106}, 105301 (2011). 
\bibitem{Pietila} V. Pietil\"{a}, Phys. Rev. A \textbf{86}, 023608 (2012).
\bibitem{Klimin} S. N. Klimin, J. Tempere, and J. T. Devreese, New J. Phys. \textbf{14}, 103044 (2012).
\bibitem{Ngampruetikorn} V. Ngampruetikorn, J. Levinsen, and M. M. Parish,
Phys. Rev. Lett. \textbf{111}, 265301 (2013). 
\bibitem{Watanabe2} R. Watanabe, S. Tsuchiya, and Y. Ohashi,
Phys. Rev. A \textbf{88}, 013637 (2013).

\bibitem{Bauer} M. Bauer, M. M. Parish, and T. Enss,
Phys. Rev. Lett. \textbf{112}, 135302 (2014).
\bibitem{Marsiglio} F. Marsiglio, P. Pieri, A. Perali, F. Palestini, and G. C. Strinati,
Phys. Rev. B \textbf{91}, 054509 (2015).

\bibitem{Murthy} P. A. Murthy, M. Neidig, R. Klemt, L. Bayha, I. Boettcher, T. Enss,
M. Holten, G. Z\"{u}rn, P. M. Preiss, and S. Jochim,
Science \textbf{359}, 6374 (2018).
\bibitem{Matsumoto} M. Matsumoto, R. Hanai, D. Inotani, and Y. Ohashi,
J. Phys. Soc. Jpn. \textbf{87}. 014301 (2018).

\bibitem{Mulkerin} B. C. Mulkerin, X.-J. Liu, and H. Hu,
arXiv:2003.06095

\bibitem{Guan} X.-W. Guan, M. T. Batchelor, and C. Lee,
Rev. Mod. Phys. \textbf{85}, 1633 (2013).

\bibitem{Imambekov} A. Imambekov, T. L. Schmidt, L. I. Glazman,
Rev Mod. Phys. \textbf{84}, 1253 (2012).

\bibitem{DrutQMC}  M. D. Hoffman, P. D. Javernick, A. C. Loheac, W. J. Porter,
E. R. Anderson, and J. E. Drut,
Phys. Rev. A \textbf{91}, 033618 (2015).
\bibitem{Drutimagmu} A. C. Loheac, J. Braun, J. E. Drut, and D. Roscher,
Phys. Rev. A \textbf{92}, 063609 (2015).
\bibitem{Rammelmuller2017} L. Rammelm\"{u}ller, W. J. Porter, J. E. Drut, and J. Braun,
Phys. Rev. D \textbf{96}, 094506 (2017).
\bibitem{Shill} C. R. Shill and J. E. Drut, Phys. Rev. A \textbf{98}, 053615 (2018).
\bibitem{Alexandru} A. Alexandru, P. F. Bedaque, and N. C. Warrington,
Phys. Rev. D \textbf{98}, 054514 (2018).



\bibitem{FF} P. Fulde and R. A. Ferrell, Phys. Rev. \textbf{135}, A550 (1964).
\bibitem{LO} A. Larkin and Y. Ovchinnikov, Zh.Eksp.Teor.Fiz. \textbf{47}, 1136 (1964).

\bibitem{Hu2007} H. Hu, X.-J. Liu, P. D. Drummond,
Phys. Rev. Lett. \textbf{98}, 070403 (2007).
\bibitem{Parish2007} M. M. Parish, S. K. Baur, E. J. Mueller, and D. A. Huse,
Phys. Rev. Lett. \textbf{99}, 250403 (2007).
\bibitem{Orso} G. Orso,
Phys. Rev. Lett. \textbf{98}, 070402 (2007).
\bibitem{Zhao} E. Zhao and W. V. Liu, Phys. Rev. A \textbf{78}, 063605 (2008).
\bibitem{Reza} M. Reza Bakhtiari, M. J. Leskinen, and P. T\"{o}rm\"{a},
Phys. Rev. Lett. \textbf{101}, 120404 (2009).
\bibitem{Liao} Y.-A. Liao, A. S. C. Rittner, T. Paprotta, W. Li, G. B. Partridge,
R. G. Hulet, S. K. Baur, and E. J. Mueller, Nature \textbf{467}, 567 (2010). 
\bibitem{Rammelmuller2020} L. Rammelm\"{u}ller, J. E. Drut, and J. Braun,
arXiv:2003.06853

\bibitem{Nakano} E. Nakano and T. Tatsumi, 
Phys. Rev. D \textbf{71}, 114006 (2005).
\bibitem{Fukushima} K. Fukushima and T. Hatsuda,
Rep. Prog. Phys. \textbf{74}, 014001 (2011).
\bibitem{Buballa} M. Buballa and S. Carignano,
Prog. Part. Nucl. Phys. \textbf{81}, 39 (2015).


\bibitem{DrutCL} A. C. Loheac, J. Braun, and J. E. Drut,
Phys. Rev. D \textbf{98}, 054507 (2018).
\bibitem{Berger} C. E. Berger, L. Rammelm\"{u}ller, A. C. Loheac,
F. Ehmann, J. Braun, and J. E. Drut, arXiv:1907.10183

\bibitem{WernerCastin} F. Werner and Y. Castin, Phys. Rev. A \textbf{86}, 013626 (2012).

\bibitem{Jin} M. Jin, M. Urban, and P. Schuck,
Phys. Rev. C \textbf{82}, 024911 (2010).
\bibitem{Ramanan} S. Ramanan and M. Urban, Phys. Rev. C \textbf{88}, 054325 (2013).
\bibitem{Pieter} P. van Wyk, H. Tajima, D. Inotani, A. Ohnishi, and Y. Ohashi,
Phys. Rev. A \textbf{97}, 013601 (2018).
\bibitem{Tajima2019} H. Tajima, T. Hatsuda, P. van Wyk, and Y. Ohashi,
Sci. Rep.  \textbf{9}, 18477 (2019).



\bibitem{Doggen} E. V. H. Doggen and J. J. Kinnunen,
Phys. Rev. Lett. \textbf{111}, 025302 (2013).

\bibitem{He} W.-B. He, Y.-Y. Chen, S. Zhang, and X.-W. Guan,
Phys. Rev. A \textbf{94}, 031604(R) (2016).

\bibitem{Voit} J. Voit, Eur. Phys. J. B. \textbf{5}, 505 (1998).
\bibitem{Orignac} E. Orignac and D. Poilblanc, Phys. Rev. B \textbf{68}, 052504 (2003).

\bibitem{Fuchs} J. N. Fuchs, A. Recati, and W. Zwerger,
Phys. Rev. Lett. \textbf{93}, 090408 (2004).

\bibitem{Rothe} H. J. Rothe, {\it Lattice Gauge Theories}, (World Scientific, 2012),
and its references.

\bibitem{Mermin} N. D. Mermin and H. Wagner,
Phys. Rev. Lett. \textbf{17}, 1133 (1966).
\bibitem{Hohenberg} P. Hohenberg, Phys. Rev. \textbf{158}, 383 (1967). 

\bibitem{FW} A. L. Fetter and J. D. Walecka, 
{\it Quantum Theory of Many-Particle Systems} (Dover Publication, Inc., Mineola, New York, 2003).

\bibitem{Bartosch} L. Bartosch, Ann. Phys. \textbf{10}, 799 (2001).

\bibitem{Giamarchi} A.-M. Visuri, M. Lebrat, S. H\"{a}usler, L. Corman, and T. Giamarchi,
Phys. Rev. Research \textbf{2}, 023062 (2020).

\bibitem{Zinner} A. S. Dehkharghani, A. G. Volosniev, and N. T. Zinner,
Phys. Rev. Lett. \textbf{121}, 080405 (2018).
\bibitem{Mistakidis} S. I. Mistakidis, G. C. Katsimiga, G. M. Koutentakis, and P. Schmelcher,
New J. Phys. \textbf{21}, 043032 (2019).


\end{thebibliography}
\end{document}